\documentclass[conference]{IEEEtran}

\makeatletter
\def\ps@IEEEtitlepagestyle{%
  \def\@oddfoot{\mycopyrightnotice}%
  \def\@evenfoot{}%
}
\def\mycopyrightnotice{%
  {\footnotesize XXX-X-XXXX-XXXX-X/XX/\$XX.00~\copyright~20XX IEEE\hfill}
  \gdef\mycopyrightnotice{}
}

\usepackage{booktabs}
\usepackage{url}
\usepackage{blindtext}
\usepackage{eso-pic}
\IEEEoverridecommandlockouts
\usepackage{cite}
\usepackage{amsmath,amssymb,amsfonts}
\usepackage{algorithmic}
\usepackage{graphicx}
\usepackage{textcomp}
\usepackage{xcolor}
\usepackage{pifont}  
\def\BibTeX{{\rm B\kern-.05em{\sc i\kern-.025em b}\kern-.08em
    T\kern-.1667em\lower.7ex\hbox{E}\kern-.125emX}}
    
\usepackage{eso-pic}

\begin{document}
\title{\vspace*{1cm} A Reverse Mamba Attention Network for Pathological Liver Segmentation\\
\thanks{This project is funded by NIH grants: R01-CA246704, R01-CA240639, U01-CA268808, and R01-HL171376.}
}

\author{\IEEEauthorblockN{1\textsuperscript{st} Jun Zeng}
\IEEEauthorblockA{\textit{School of Software Engineering} \\
\textit{Chongqing University of Posts and Telecommunications}\\
Chongqing, China \\
zeng.cqupt@gmail.com}
\and
\IEEEauthorblockN{2\textsuperscript{nd} Debesh Jha}
\IEEEauthorblockA{\textit{ Department of Computer Science} \\
\textit{University of South Dakota}\\
Vermillion SD, USA \\
debesh.jha@usd.edu}
\and
\IEEEauthorblockN{3\textsuperscript{rd} Ertugrul Aktas}
\IEEEauthorblockA{\textit{Machine \& Hybrid Intelligence Lab.} \\
\textit{Northwestern University}\\
Chicago, IL, USA \\
h.ertugrulaktas@gmail.com}
\and
\IEEEauthorblockN{4\textsuperscript{th} Elif Keles}
\IEEEauthorblockA{\textit{Machine \& Hybrid Intelligence Lab.} \\
\textit{Northwestern University}\\
Chicago, IL, USA \\
elifkeles.dr@gmail.com}
\and
\IEEEauthorblockN{5\textsuperscript{th} Alpay Medetalibeyoglu}
\IEEEauthorblockA{\textit{Machine \& Hybrid Intelligence Lab.} \\
\textit{Northwestern University}\\
Chicago, IL, USA \\
alibeyoglualpay@gmail.com}
\and
\IEEEauthorblockN{6\textsuperscript{th} Matthew Antalek}
\IEEEauthorblockA{\textit{Department of Interventional Radiology} \\
\textit{Northwestern University}\\
Chicago, IL, USA \\
matthew.antalek@nm.org}
\and
\IEEEauthorblockN{7\textsuperscript{th} Robert Lewandowski}
\IEEEauthorblockA{\textit{Department of Interventional Radiology} \\
\textit{Northwestern University}\\
Chicago, IL, USA \\
r-lewandowski@northwestern.edu}
\and
\IEEEauthorblockN{8\textsuperscript{th} Daniela Ladner}
\IEEEauthorblockA{\textit{Department of Surgery} \\
\textit{Northwestern University}\\
Chicago, IL, USA \\
dladner@nm.org}
\and
\IEEEauthorblockN{9\textsuperscript{th} Amir A. Borhani}
\IEEEauthorblockA{\textit{Department of Radiology} \\
\textit{Northwestern University}\\
Chicago, IL, USA \\
amir.borhani@nm.org}
\and
\IEEEauthorblockN{10\textsuperscript{th} Gorkem Durak}
\IEEEauthorblockA{\textit{Machine \& Hybrid Intelligence Lab.} \\
\textit{Northwestern University}\\
Chicago, IL, USA \\
gorkem.durak@northwestern.edu}
\and
\IEEEauthorblockN{11\textsuperscript{th} Ulas Bagci}
\IEEEauthorblockA{\textit{Machine \& Hybrid Intelligence Lab.} \\
\textit{Northwestern University}\\
Chicago, IL, USA \\
ulas.bagci@northwestern.edu}
}

\maketitle
\begin{abstract}
We present RMA-Mamba, a novel architecture that advances the capabilities of vision state space models through a specialized reverse mamba attention module (RMA). The key innovation lies in RMA-Mamba's ability to capture long-range dependencies while maintaining precise local feature representation through its hierarchical processing pipeline. By integrating Vision Mamba (VMamba)'s efficient sequence modeling with RMA's targeted feature refinement, our architecture achieves superior feature learning across multiple scales. This dual-mechanism approach enables robust handling of complex morphological patterns while maintaining computational efficiency. We demonstrate RMA-Mamba's effectiveness in the challenging domain of pathological liver segmentation (from both CT and MRI), where traditional segmentation approaches often fail due to tissue variations. When evaluated on a newly introduced cirrhotic liver dataset (CirrMRI600+) of T2-weighted MRI scans, RMA-Mamba achieves the state-of-the-art performance with a Dice coefficient of 92.08\%, mean IoU of 87.36\%, and recall of 92.96\%. The architecture's generalizability is further validated on the cancerous liver segmentation from CT scans (LiTS: Liver Tumor Segmentation dataset), yielding a Dice score of 92.9\% and mIoU of 88.99\%. Our code is available for public: {\color{blue}{\url{https://github.com/JunZengz/RMAMamba}}}.
\end{abstract}


\section{Introduction}
\textbf{Technical motivation.} The Transformer architecture~\cite{vaswani2017attention} significantly advanced sequence modeling by employing self-attention mechanisms to capture long-range dependencies without relying on recurrence. Despite their success, Transformers face limitations in handling long sequences due to the quadratic complexity of the self-attention operation with respect to sequence length. To address this challenge, recent studies have explored the integration of State Space Models (SSMs) into Transformer architectures, leading to the development of Mamba-style architectures. These architectures converge the strengths of Transformers and SSMs by replacing the standard self-attention with linear recurrent layers derived from state space formulations. Specifically, they utilize parameterizations that enable efficient computation of long-range dependencies with linear complexity, such as the Structured State Space Sequence model (S4). This convergence allows Mamba architectures to maintain the expressive power of Transformers while significantly improving scalability and efficiency in modeling long sequences. The evolution from pure self-attention mechanisms to incorporating SSMs represents a significant step forward in designing architectures that are both computationally efficient and capable of capturing complex dependencies in high-resolution computer vision tasks~\cite{gu2023mamba}. 

The adaptation of SSMs to computer vision was pioneered by S4ND~\cite{nguyen2022s4nd}, which extended the model to handle multi-dimensional data and surpassed Vision Transformer benchmarks on ImageNet-1k. VMamba~\cite{liu2024vmamba} further advanced this approach by introducing a multi-route scanning mechanism and replacing the traditional S6 block with SS2D, creating the innovative VSS block for robust visual feature extraction. Recent implementations of VSS blocks in medical image segmentation have shown promising results~\cite{ruan2024vmunet, wang2024mambaunet, liu2024swin}. Building on these advances, we propose RMA-Mamba, which leverages and enhances the VSS block's capacity to simultaneously capture local detail and global context - a crucial capability for accurate pathological liver segmentation. Our architecture introduces novel mechanisms to optimize the VSS block specifically for medical imaging applications, addressing the unique challenges of tissue boundary delineation and pathological feature detection.

\textbf{Clinical motivation.} The liver, vital for metabolism and digestion, is increasingly affected by cirrhosis~\cite{plosone} and represents a critical site for both primary and metastatic cancer development. As the sixth most common location for primary malignancies, it also frequently harbors metastases from other abdominal organs, particularly the colon and pancreas~\cite{donne2023liver}. Early detection of liver pathologies significantly impacts patient survival rates~\cite{women} and treatment options. While modern imaging modalities like computed tomography (CT) and magnetic resonance imaging (MRI) provide detailed visualization of hepatic tissue, the current clinical workflow still relies heavily on manual or semi-automated slice-by-slice contour delineation by radiologists. This process is not only time-intensive but also subject to inter-observer variability and expertise-dependent accuracy. The development of robust computer-assisted segmentation algorithms for pathological liver tissue addresses these clinical challenges by offering consistent, automated analysis of medical images. Such tools promise to reduce manual annotation burden while potentially improving diagnostic accuracy through standardized quantitative assessment.

While liver segmentation has seen remarkable progress over recent decades~\cite{khoshkhabar2023automatic, zheng2022automatic, gorade2024rethinking}, the specific challenge of cirrhotic liver segmentation from MRI remains underexplored, and cancerous liver segmentation (from CT and MRI) have significant limitations. This gap is particularly concerning given that early-stage cirrhosis often goes undetected in clinical practice, leading to disease progression toward decompensated cirrhosis or hepatocellular carcinoma (HCC). The increasing mortality rates associated with this progression underscore the urgent need for robust segmentation tools that can later be used in a computer aided diagnosis and quantification classifier to reliably identify cirrhotic changes across diverse patient populations and imaging protocols. The recent introduction of the CirrMRI600+ dataset~\cite{Jha_Bagci_2024} represents a significant step toward addressing this unmet clinical need. This comprehensive dataset enables the development and validation of advanced segmentation algorithms capable of handling the subtle tissue changes characteristic of early cirrhosis, while maintaining robustness across different MRI scanners and patient demographics.

\textbf{What do we propose?} 
\begin{itemize}
\item   \textbf{RMA-Mamba: a new architecture:} 
We introduce RMA-Mamba, a novel architecture that leverages VMamba's hierarchical feature extraction capabilities enhanced by our proposed Reverse Mamba Attention module. This design marks the first application of state space models to pathological liver segmentation from MRI and CT scans, establishing a new paradigm for medical image analysis tasks requiring fine-grained tissue differentiation.

\item \textbf{Reverse Mamba Attention (RMA) Module:} Our key innovation lies in the RMA module, which implements a progressive feature integration strategy during decoder operations. RMA's unique approach to feature enrichment enables simultaneous capture of microscopic tissue variations and macroscopic anatomical context, addressing the dual challenge of local detail preservation and global structure understanding of liver anatomy.

\item  \textbf{{Rich benchmarking results}}: Extensive benchmarking against eight state-of-the-art methods demonstrates RMA-Mamba's superior performance, achieving a dice coefficient of 92.08\%, mean IoU of 87.36\%, and recall of 92.96\% on CirrMRI600+. These results validate our architectural innovations and establish new performance standards for cirrhotic and cancerous liver segmentation. The model's effectiveness is further confirmed through rigorous evaluation on the LiTS dataset, demonstrating robust generalization across different imaging protocols.
\end{itemize}

\section{Related Work}
\subsection{State Space Model}
SSMs have emerged as powerful architectures for sequence modeling~\cite{fu2022hungry}, with each iteration addressing key computational and performance challenges. The Linear State Space Layer (LSSL)~\cite{gu2021efficiently} pioneered the use of HIPPO (High-order Polynomial Projection Operators) matrices and structured matrices for capturing long-range dependencies. However, its computational complexity limited practical applications for extended sequences.

S4~\cite{gu2021efficiently} marked a significant advancement by introducing novel parameterization techniques that transform HIPPO matrices into diagonal plus low-rank representations. This innovation, coupled with implicit diagonalization, substantially reduced computational overhead for long sequence processing. Building on this foundation, S5~\cite{smith2022simplified} further streamlined computations through diagonal approximation and parallel scanning mechanisms, achieving both improved efficiency and performance.

Mamba~\cite{gu2023mamba} represents the current state-of-the-art through its S6 module, which revolutionizes the approach by converting fixed inputs into variable function forms. Its selective scanning mechanism effectively replaces traditional convolutions, enabling intelligent filtering of information flow. This architectural innovation allows Mamba to dynamically focus on relevant features while discarding redundant information.

\subsection{Liver segmentation}
Deep learning has transformed medical image segmentation, initially with UNet \cite{ronneberger2015u}, later with nnUnet, establishing a foundational encoder-decoder architecture that efficiently combines high-resolution features with contextual information through skip connections. These architectures have spawned numerous innovations in liver segmentation~\cite{li2020attention, jin2020ra, kushnure2021ms, WANG2021sar-unet, gao2021asu}, each addressing specific aspects of the segmentation challenge.

Attention mechanisms have emerged as a key enhancement to the basic UNet architecture. AttentionUNet++~\cite{li2020attention} introduced attention gates with dense connections, while Jin et al. developed a dual-branch approach combining attention residual learning with feature enhancement through a sophisticated trunk-and-mask architecture. SAR-UNet~\cite{WANG2021sar-unet} further refined this approach by incorporating squeeze-and-excitation blocks for adaptive feature enhancement and atrous spatial pyramid pooling for multi-scale context capture, while addressing gradient flow through residual connections.

Recent architectures have focused on multi-scale feature learning and boundary refinement. Xie et al.~\cite{xie2022mci} proposed a multi-scale context extraction module with external attention and boundary correction mechanisms. Liu et al.~\cite{liu2023gcha} introduced a hybrid attention approach combining global dependencies with local feature focus. 

\begin{figure*} [!t]
    \centering
    \includegraphics[width=0.65\textwidth]{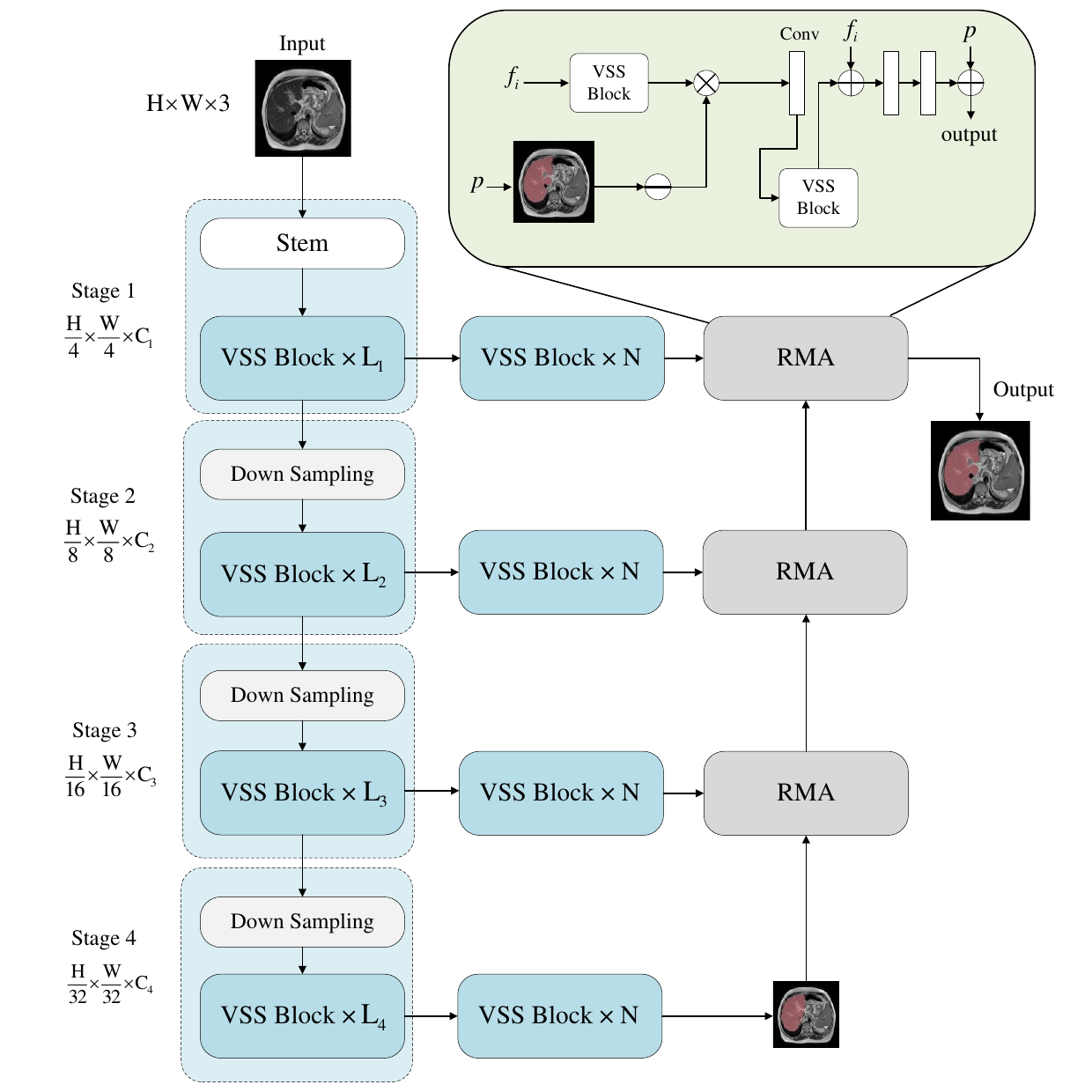}
    \caption{Overview of the proposed \textit{RMA-Mamba} architecture.}
    \label{fig:RMAMamba}
\end{figure*}

\section{Method}{\label{section:method}}
In this section, we first provide an overview of the proposed RMA-Mamba architecture. Next, we discuss the specifics of the encoder used in this study. Following that, we present a detailed explanation of the VSS and RMA modules. Finally, we introduce the loss function in detail.

\subsection{Overview of the architecture}
Figure~\ref{fig:RMAMamba} shows the overview of the proposed RMA-Mamba architecture. As illustrated, RMA-Mamba consists of three key components: the VMamba encoder, additional VSS blocks, and the RMA module. The synergy of these components significantly enhances the model's learning capability and comprehensive segmentation performance. In the encoding process, the pre-trained VMamba backbone extracts four different levels of features $\frac{H}{2^{i+1}} \times \frac{W}{2^{i+1}} \times C_{i}$ from the input liver image, where $C_{i} \in \left\{96, 192, 384, 768 \right\}$ and $i \in \left\{1, 2, 3, 4 \right\}$. Additional VSS blocks are then applied to refine the semantic representation of these features, Where N denotes the number of VSS blocks. Finally, the refined features are passed through the RMA to produce the final segmentation result.

Moreover, we also introduce RMA-Mamba-T and RMA-Mamba-S submodules, designed to meet varying requirements in liver segmentation tasks. Specifically, RMA-Mamba-T employs pre-trained VMamba-Tiny as the backbone, with the number of additional VSS blocks set to 0. This variant employs fewer parameters and achieves faster inference speed. RMA-Mamba-S, on the other hand, utilizes pre-trained VMamba-Small as the backbone, with one additional VSS block, resulting in more parameters but improved segmentation capabilities.

\subsection{Encoder}
The proposed architecture adopts VMamba as the encoder backbone, capitalizing on its efficient inference capabilities and high throughput characteristics. VMamba~\cite{liu2024vmamba}, built upon the selective SSM paradigm, offers \textit{linear computational complexity} while maintaining strong scalability for vision tasks. The encoder's architecture progressively processes input images through multiple stages:
\begin{itemize}
    \item A stem module that transforms the input image into initial feature maps of dimension $\frac{H}{4} \times \frac{W}{4}$,
    \item Four sequential processing stages, where all but the first incorporate VSS blocks preceded by down-sampling layers,
    \item  Multi-scale feature representations generated at each stage, crucial for comprehensive visual understanding.
\end{itemize}

VMamba offers three model variants (Tiny, Small, and Base) to balance computational requirements with performance. Our empirical analysis led us to implement VMamba-Tiny and VMamba-Small variants, as they achieve optimal performance-efficiency trade-offs for pathological liver segmentation. The larger VMamba-Base variant, despite its increased parameter count, did not demonstrate commensurate performance gains to justify its computational overhead.

\subsection{VSS block}
While traditional visual data processing relies heavily on spatial relationships, unidirectional scanning mechanisms often fail to capture comprehensive dependencies across image regions. Although Transformers~\cite{vaswani2017attention} successfully address this limitation through self-attention mechanisms that capture long-range dependencies, they present computational challenges at scale.

The 2D-Selective-Scan module (SS2D)~\cite{gu2023mamba} offers an elegant solution by implementing four distinct scanning routes across feature maps, effectively capturing global receptive fields while maintaining computational efficiency. This innovation evolved from replacing Mamba's S6 module with SS2D in the vanilla VSS block design. VMamba further refined this approach by adopting Transformer-like architectural principles, resulting in streamlined VSS blocks that combine SS2D with Feed-Forward Network (FFN) layers in a single processing branch.

Our network implements a multi-stage feature refinement process:
\begin{enumerate}
    \item Hierarchical features from the backbone undergo enhancement through VSS blocks.
    \item These enhanced features are processed by $3\times3$ convolutional layers to normalize channel dimensions to 32.
    \item The first three feature levels are directed through their respective reverse mamba attention modules for further refinement.
    \item The final feature level undergoes dimensionality reduction via a $1\times1$ convolution followed by sigmoid activation, generating an initial binary segmentation map.
\end{enumerate}

\subsection{Reverse Mamba Attention Module}
The reverse attention mechanism has demonstrated remarkable success in salient object detection~\cite{zhang2021bilateral}, particularly in boundary delineation and feature refinement. This success was notably exemplified in PraNet~\cite{fan2020pranet}, where reverse attention effectively enhanced boundary detection and tissue differentiation, leading to improved segmentation accuracy through targeted error correction. Building on this foundation, RTA-Former~\cite{li2024rta} advanced the concept by introducing transformer-based reverse attention, demonstrating superior edge segmentation capabilities through sophisticated feature extraction. These successes in boundary refinement and feature enhancement inspired our novel approach: leveraging the computational efficiency of VSS blocks within a reverse attention framework for pathological liver.

Our proposed RMA module represents a significant advancement in this trajectory, specifically designed to address the unique challenges of pathological liver segmentation. By combining the selective scanning capabilities of VSS blocks with reverse attention principles, RMA enables precise tissue boundary delineation while maintaining computational efficiency.

As illustrated in Figure~\ref{fig:RMAMamba}, RMA obtains complementary information from the adjacent upper stage and integrates it into the previous segmentation map by an addition operation. First, the initial binary segmentation map $P_{i+1}$ and shallow feature $f_{i}$ are processed by the reverse operation to obtain the attention map. Next, we perform element-wise multiplication to obtain reverse attention features $R_{i}$. This process is formulated as follows:
    \begin{equation} \label{revserse}
	R_{i} = \circleddash(P_{i+1}) \odot \delta(f_{i}),
    \end{equation}
where $\circleddash$ denotes a reverse operation, where the input is subtracted from the matrix $\mathbf{E}$, with all elements equal to $1$. $\delta(\cdot)$ denotes processing by a VSS block, and $\odot$ denotes element-wise multiplication. To further supplement the information, we reintroduce $f_{i}$ and perform an element-wise addition, followed by two convolutional layers to generate the feature map $p_{2}$. Finally, we perform an addition operation between $p$ and $p_{2}$ to obtain the final output.

\subsection{Loss function}
In medical image segmentation tasks with deep learning, Binary Cross-Entropy and Dice losses are commonly employed to improve the precision of delineations. During the training phase, we combine these two loss functions to effectively guide and supervise the outputs of our model, and are defined conventionally as:
\begin{equation}
\begin{aligned}
    L_{\text{BCE}} &= -\frac{1}{n} \sum_{i=1}^{n} \left[ y_i \log(\hat{y}_i) + (1 - y_i) \log(1 - \hat{y}_i) \right],    \mbox{ and } 
\end{aligned}
\end{equation}
\begin{equation}
\begin{aligned}
 L_{\text{Dice}} &= 1 - \frac{2 \sum_{i=1}^{n} y_i \hat{y}_i}{\sum_{i=1}^{n} y_i + \sum_{i=1}^{n} \hat{y}_i},
\end{aligned}
\end{equation}where $n$ represents the total number of samples, $y_i$ represents the true label, $\hat{y}_i$ denotes the predicted value.

\begin{table*}[t!]
\centering
\scriptsize
\caption{Model performance on the CirrMRI600+ dataset.}  
 \begin{tabular} {c|c|c|c|c|c|c|c}
\toprule

\textbf{{Method}}  & \textbf{{Publication}}  &\textbf{Dice (\%)} & \textbf{mIoU (\%)}  &\textbf{Recall (\%)} & \textbf{Precision (\%)} & \textbf{F2 (\%)} & \textbf{{HD}}  \\ 
\hline

U-Net~\cite{ronneberger2015u}  &MICCAI 2015 &89.40 &84.10 &90.95 &91.66  &90.01 &3.52  \\





UNeXt~\cite{valanarasu2022unext} &MICCAI 2022	&83.83 &76.88 &85.57 &88.46  &84.46 &4.03  \\

TransNetR~\cite{jha2023transnetr} & MIDL 2023 &90.35  &85.05  &91.51  &92.10 &90.78 &3.50  \\

TransResUNet~\cite{tomar2022transresu} & IEEE CMBS 2023 &91.47 &86.52  &92.70   &92.60 &91.95 &3.42  \\


VM-UNet~\cite{ruan2024vmunet} & arXiv 2024 &90.34 &85.08 &92.31 &91.51 &91.18 &3.52 \\

VM-UNetV2~\cite{zhang2024vm} & ISBRA 2024 &89.42 &83.74 &90.44 &91.37 &89.73 &3.55 \\

UltraLight VM-UNet~\cite{wu2024ultralight} & arXiv 2024 &72.03 &63.18 &75.21 &77.59 &73.23 &4.77   \\

Mamba-UNet~\cite{wang2024mambaunet} & arXiv 2024 &90.66  &85.57 &92.63 &91.73 &91.55 &3.49\\

\textbf{RMAMamba-T} & Proposed &\textcolor{black}{91.17} &\textcolor{black}{86.16} &92.34 & 92.48 &91.65	&3.50
\\

\textbf{RMAMamba-S} & Proposed &\textcolor{black}{\textbf{92.08}} &\textcolor{black}{\textbf{\textbf{87.36}}} &\textbf{92.96} & \textbf{93.30} &\textbf{92.42}	&\textbf{3.39}
\\
\bottomrule
\end{tabular}
\label{tab:segmentationcirrmri600plus}
\end{table*}

\begin{figure*} [!t]
    \centering
    \includegraphics[width=0.8\textwidth]{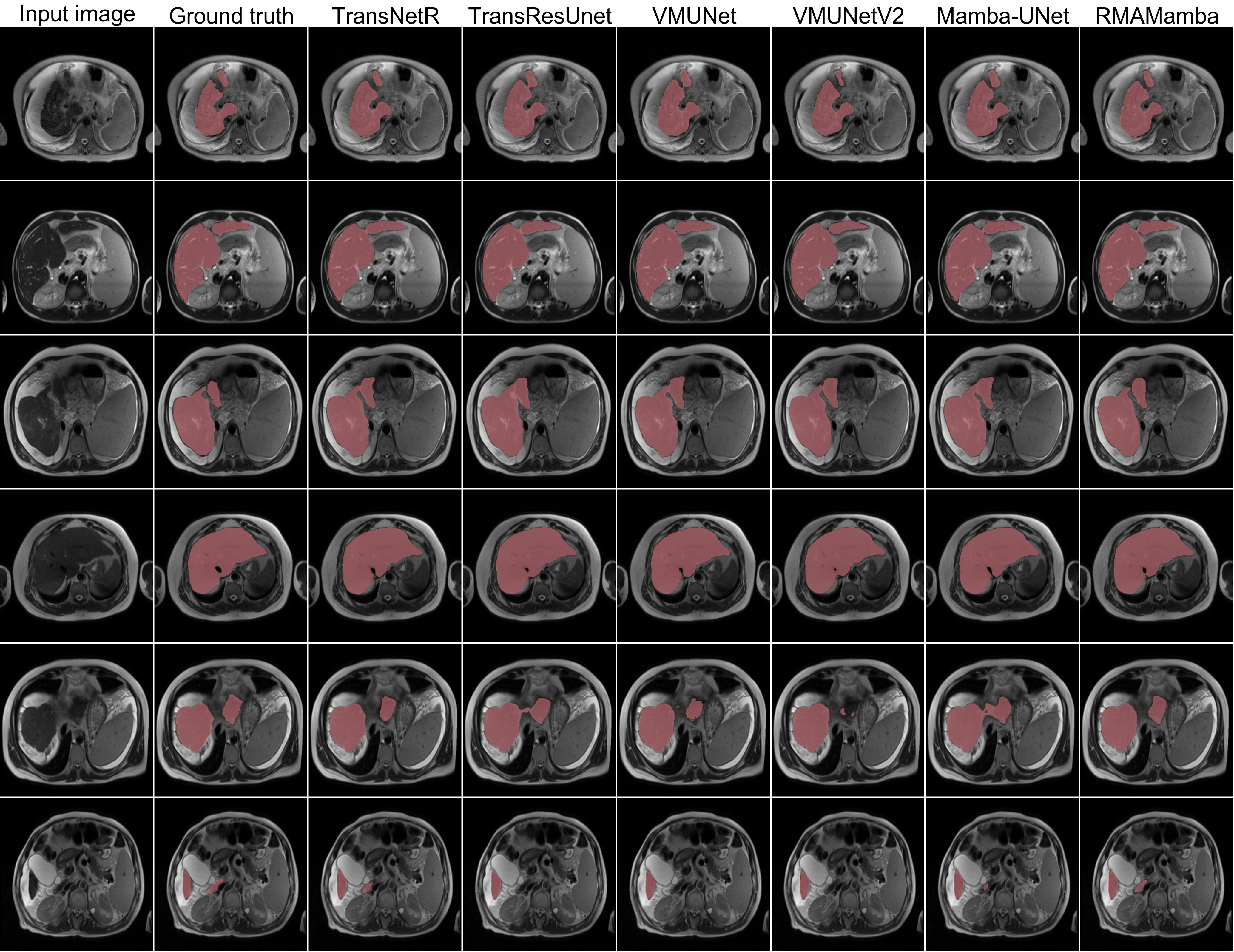}
    \caption{Qualitative results of different methods on the CirrMRI600+ dataset.}
    \label{fig:CirrMRI600+}
\end{figure*}

\begin{table*}[t!]
\scriptsize
\centering
\caption{Model performance on the LiTS dataset.}  
 \begin{tabular}{c|c|c|c|c|c|c|c}
\toprule

\textbf{{Method}}  & \textbf{{Publication}}  &\textbf{Dice (\%)} & \textbf{mIoU (\%)}  &\textbf{Recall (\%)} & \textbf{Precision (\%)} & \textbf{F2 (\%)} & \textbf{{HD}}  \\ 
\hline

U-Net~\cite{ronneberger2015u}  &MICCAI 2015 &90.79 &86.31 &92.10 &92.49  &91.37 &2.97  \\





UNeXt~\cite{valanarasu2022unext} &MICCAI 2022	&90.13 &85.46 &89.87 &94.13  &89.78 &2.99  \\

TransNetR~\cite{jha2023transnetr} & MIDL 2023 &91.88 &87.66 &91.32 &95.39  &91.41 &2.89  \\ 

TransResUNet~\cite{tomar2022transresu} & IEEE CMBS 2023 &92.58 &88.42 &92.41 &95.01  &92.34 &2.86  \\ 


VM-UNet~\cite{ruan2024vmunet} & arXiv 2024 &90.82 &86.34 &90.45 &94.16  &90.39 &2.94  \\ 

VM-UNetV2~\cite{zhang2024vm} & ISBRA 2024 &92.05 &87.95 &91.98 &94.86  &91.84 &2.87  \\ 

UltraLight VM-UNet~\cite{wu2024ultralight} & arXiv 2024 &85.77 &79.77 &85.65 &89.76  &85.51 &3.40  \\ 

Mamba-UNet~\cite{wang2024mambaunet} & arXiv 2024 &91.88 &87.77 &92.17 &94.42  &91.82 &2.89  \\

\textbf{RMAMamba-T} & Proposed &\textbf{92.94} &\textbf{88.99} &92.34 &\textbf{95.59}  &92.44 &\textbf{2.83}  \\ 

\textbf{RMAMamba-S} & Proposed &92.74 &88.92 &\textbf{92.91} &94.92  &\textbf{92.67} &2.86  \\

\bottomrule
\end{tabular}
\label{tab:segmentationlitS}
\end{table*}

\begin{figure*} [!t]
    \centering
    \scriptsize
    \includegraphics[width=0.8\textwidth]{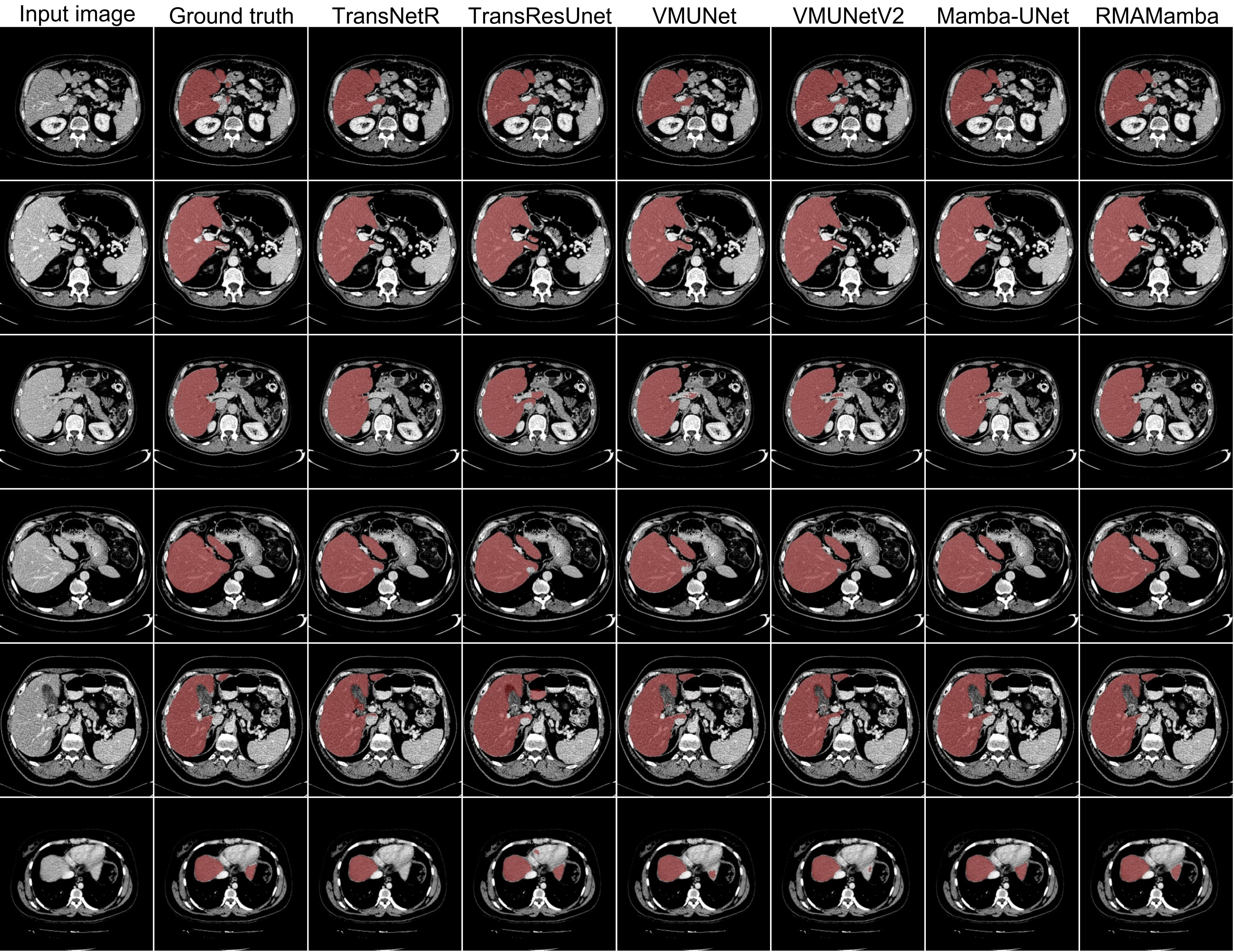}
    \caption{Qualitative results of different methods on the LiTS dataset.}
    \label{fig:LiTS}
\end{figure*}

\section{Experiments and Results}
\label{sec:experiments}

\subsection{Datasets}
\paragraph{\textbf{CirrMRI600+ dataset:}}
The CirrMRI600+\cite{Jha_Bagci_2024} dataset is a single-center, multivendor, multiplanar, and multiphase dataset specifically designed for cirrhotic liver research. It includes 628 high-resolution abdominal MRI scans from 339 patients, with 310 T1-weighted (T1W) and 318 T2-weighted (T2W) volumetric scans with corresponding segmentation masks.
All experiments are conducted using this predefined dataset split.
\paragraph{\textbf{LiTS dataset:}}
LiTS~\cite{bilic2023liver} is a multicenter abdominal CT dataset (201 scans) with expert annotations for liver and tumor segmentation.  The training set (130 scans) was used to generate healthy liver masks, which were then divided into training (11684 slices), validation (2745 slices), and test (4734 slices) sets. Slice count variations in validation and test sets reflect patient-specific CT slice numbers.

\subsection{Implementation}
In our development environment, we implemented RMAMamba using Ubuntu 20.04, Python 3.8, PyTorch 1.13, and CUDA 11.7. All experiments were conducted using the PyTorch framework on a single NVIDIA RTX A10 GPU with 24GB of memory. For RMAMamba-T and RMAMamba-S, we initialize the encoder weights with VMamba-T and VMamba-S respectively, both pre-trained on ImageNet-1k. During training, the images in the T2-2D dataset were resized to 256 × 256, with data augmentation techniques including random rotation, vertical flip, and horizontal flip. We utilized a combination of binary cross-entropy and dice loss for the supervision of outputs. We trained our model using the Adam optimizer with a learning rate of 1e-4. \textit{ReduceLROnPlateau} is utilized as the scheduler with a patience of 5 epochs. The models were trained for a maximum of 500 epochs, with an early stopping patience of 50. The batch size was initiated to 16.

\begin{table*}[t!]
\centering
\caption{Ablation study on the CirrMRI600+ dataset.}  
 \begin{tabular} {c|c|ccc|c|c|c|c}
\toprule

{\textbf{{Exp.}}} &
{\textbf{{Backbone}}} &
{\textbf{{Setting}}} &
{\textbf{{RA}}} &
{\textbf{{RMA}}} &
\textbf{Dice (\%)} & \textbf{mIoU (\%)}  &
\textbf{F2 (\%)} & 
\textbf{{HD}}  \\
\hline

\#1 &VMamba-Tiny &N=0  &\ding{51} &\ding{55} &90.86 &85.63 &91.50  &3.52  \\ 
\#2 &VMamba-Small &N=0 &\ding{51} &\ding{55}  &91.08  &86.16 &91.09 &3.46   \\
\#3 &VMamba-Small &N=0 &\ding{55} &\ding{51} &91.45 &86.31 &92.05  &3.47  \\ 
\#4 &VMamba-Small &N=1 &\ding{51} &\ding{55} &91.92 &87.05 &92.34  &3.42  \\ 
\#5 &VMamba-Tiny &N=0 &\ding{55} &\ding{51} &91.17 &86.16 &91.65  &3.50  \\ 
\#6 &VMamba-Small &N=1 &\ding{55} &\ding{51}  &\textbf{92.08}  &\textbf{87.36} &\textbf{92.43} &\textbf{3.39}  \\ 

\bottomrule
\end{tabular}
\label{tab:ablationoncirrmri600plus}
\end{table*}

\subsection{Evaluation metrics}
For a comprehensive comparison, we utilized six common evaluation metrics: dice coefficient (Dice), mean Intersection over Union (mIoU), recall, precision, F2 score, and Hausdorff Distance (HD). Dice and IoU are reliable metrics for assessing the similarity between predicted segmentation masks and ground truth values. These two metrics are crucial for measuring overlap and consistency between the two masks and evaluating segmentation accuracy. Precision reflects the proportion of true positive samples among all the predicted positive samples. Recall measures the model's ability to identify the targeted liver tissues correctly. HD demonstrates the discrepancy between the segmentation and actual boundaries, with lower values indicating smaller differences. 

\subsection{Benchmarking}
We compared our architectures, RMAMamba-S and RMAMamba-T, with the state-of-the-art deep segmentation methods (U-Net \cite{ronneberger2015u}, UNext \cite{valanarasu2022unext}), Transformer-based method (TransResUNet \cite{tomar2022transresu}, TransNetR~\cite{jha2023transnetr}), and SSM-based segmentation methods (VM-UNet \cite{ruan2024vmunet}, VM-UNetV2 \cite{zhang2024vm}, UltraLight VM-UNet \cite{wu2024ultralight}, Mamba-UNet \cite{wang2024mambaunet}) on the CirrMRI600+ and LiTS datasets. To ensure a fair comparison, all models were trained and evaluated with identical dataset partitioning and hyperparameter settings.

\paragraph{\textbf{Results on CirrMRI600+ dataset:}}
Our extensive experimentation on the pathological liver segmentation benchmark dataset demonstrates RMA-Mamba's superior performance (Table~\ref{tab:segmentationcirrmri600plus}). RMA-Mamba-T outperforms existing CNN and Transformer-based architectures while showing marked improvement over recent SSM-based approaches like VM-UNet and Mamba-UNet. The larger variant, RMAMamba-S, achieves state-of-the-art performance with a Dice score of 92.08\%, mIoU of 87.36\%, and Hausdorff Distance of 3.39 mm.

While Transformer-based methods (TransResUnet: 92.58\%, PVTFormer: 92.48\%) and CNN-based approaches (Double-UNet: 92.37\%) demonstrate competitive performance, they incur substantial computational overhead and longer training times. RMA-Mamba achieves comparable or superior results while maintaining computational efficiency and faster training convergence. Qualitative analysis (Figure~\ref{fig:CirrMRI600+}) further validates our method's effectiveness, particularly in challenging cases. The visualization results, especially evident in rows three and five, demonstrate RMA-Mamba's superior boundary delineation and enhanced discriminative capability at tissue interfaces.

\paragraph{\textbf{Results on LiTS dataset:}}
Table~\ref{tab:segmentationlitS} presents the quantitative results of various state-of-the-art models on the LiTS dataset. We can observe that RMA-Mamba-T achieves exceptional segmentation performance, with a Dice coefficient of 92.94\%, mIoU of 88.99\%, recall of 92.34\%, precision of 95.59\%, F2 score of 92.44\%, and a HD score of 2.83 mm. RMA-Mamba-T achieved the highest Dice, surpassing PVTFormer by 0.46\%, VM-UNetV2 by 0.89\%, and ResUNet++ by 2\%. Compared to RMA-Mamba-S, RMA-Mamba-T has fewer parameters yet delivers superior performance on the LiTS dataset. Additionally, RMA-Mamba-T offers faster inference speed and lower computational cost. Figure~\ref{fig:LiTS} shows that our model provides more accurate segmentation results than other benchmarked state-of-the-art methods.

\subsection{Ablation study}
Table~\ref{tab:ablationoncirrmri600plus} summarizes the ablation study of the RMA-Mamba on the CirrMRI600+ dataset. As mentioned in Section~\ref{section:method}, RMA-Mamba consists of a VMamba backbone, additional VSS blocks beyond the backbone, and RMA modules, where N represents the number of additional VSS blocks. In Table~\ref{tab:ablationoncirrmri600plus}, RA denotes the traditional convolution-based reverse attention mechanism, originally introduced in PraNet~\cite{fan2020pranet}. 
The first group (\#1) and the second group (\#2) represent experimental setups using Vmamba-Tiny and Vmamba-Small as the network backbones, respectively, without the RMA module and additional VSS blocks. The fifth group (\#5) and the sixth group (\#6) correspond to the RMA-Mamba-T and RMA-Mamba-S architectures, respectively. 

The third group (\#3) and the fourth group (\#4) represent experimental settings where RMA-Mamba-S removed the additional VSS block and the RMA module, respectively. Results from \#1, \#2, and \#4 indicate that removing the RMA module leads to a significant drop in model performance, underscoring the crucial role of the RMA module in enhancing segmentation quality. Additionally, in \#3, removing the additional VSS block from RMA-Mamba-S also leads to a decline across multiple metrics.

Table 3 illustrates that both RMA-Mamba-S (\#6) and RMA-Mamba-T (\#5) demonstrate excellent performance on the CirrMRI600+ dataset. Therefore, we selected \#5 as RMA-Mamba-T and \#6 as RMA-Mamba-S for their complementary strengths: RMA-Mamba-T delivers the highest performance despite a larger parameter count, while RMA-Mamba-S maintains strong segmentation accuracy with fewer parameters. Depending on the requirements of different tasks, we can flexibly choose between RMA-Mamba-S and RMA-Mamba-T.

\section{Conclusion}
In this paper, we introduced RMA-Mamba, a novel architecture that leverages state space models for pathological liver segmentation from MRI and CT scans. Our approach combines VMamba's efficient hierarchical feature extraction capabilities with a specially designed Reverse Mamba Attention module to capture both fine-grained tissue details and global anatomical context. Extensive experimentation on the CirrMRI600+ and LiTS datasets demonstrates RMA-Mamba's superior performance to other state of the art methods while maintaining computational efficiency. RMA-Mamba not only advances the state-of-the-art in liver segmentation but also establishes a promising direction for applying state space models to medical image analysis tasks. While the RMA-Mamba demonstrates impressive performance in pathological liver segmentation from both MRI and CT scans, a notable limitation lies in its slice-by-slice processing approach. This approach may not fully capture the intricate 3D anatomical relationships crucial for accurate liver segmentation, particularly in the presence of complex liver pathologies. This limitation opens avenues for future research to explore extensions of RMA-Mamba or similar state space model-based architectures to incorporate 3D information. This could involve adapting the RMA module to operate in 3D or integrating the VMamba encoder into a 3D segmentation framework.


\end{document}